# Molecular adhesion assay for biopolymer systems

Jeremy A. Cribb[1], Farnaz Fazelpour[2], David A. Wollensak[1], Danielle Rice[3], Max deJong[1], David Hill[1,2,4], Richard Superfine[5]

[1]Department of Physics and Astronomy, The University of North Carolina at Chapel Hill, Chapel Hill, North Carolina 27599, USA

[2]Marsico Lung Institute, The University of North Carolina at Chapel Hill, Chapel Hill, North Carolina 27599, USA

[3]College of Engineering, North Carolina Agricultural and Technical State University

[4]Lampe Joint Department of Biomedical Engineering, The University of North Carolina at Chapel Hill, Chapel Hill, North Carolina 27599, USA

[5]Department of Applied Physical Sciences, The University of North Carolina at Chapel Hill, Chapel Hill, North Carolina 27599, USA

Author ORCID IDs:
Jeremy A Cribb: 0000-0001-9725-1209
Farnaz Fazelpour: 0000-0002-6133-5264
David Wollensak: 0009-0001-1334-1475
David Hill: 0000-0002-9270-777X
Richard Superfine: 0000-0002-2569-071X

## Abstract

Molecular adhesion plays a central role in many biological systems, yet existing methods to quantify adhesive strength often struggle to bridge the gap between single-molecule resolution and biologically relevant environments. Here, we present a scalable micromagnetic bead-based adhesion assay capable of quantifying detachment forces under physiologically meaningful conditions. Designed to probe mucoadhesion in the context of mucociliary clearance, our system applies controlled magnetic forces to ligand-coated beads adhered to functionalized substrates and tracks detachment events using high-speed microscopy and calibrated z-displacement mapping. The platform combines substrate- and bead-side surface chemistry control with high-throughput imaging and *in situ* force calibration via Stokes drag. We demonstrate the ability to distinguish sub-nanonewton to nanonewton force regimes across a range of bead-substrate pairings, including COOH-COOH, PEG-PEG, and cell culture-derived human bronchial epithelial (HBE) mucus interactions. Surface functionalization was validated via fluorescence imaging and zeta potential measurements, while detachment forces were used to estimate binding energy and infer dissociation constants. This assay enables detailed characterization of multivalent, force-sensitive adhesive interactions and offers a powerful new approach for studying bioadhesive systems, including mucus-pathogen interactions and drug delivery materials.

## Introduction

Adhesive molecular interactions regulate a host of biological and cellular processes, including tissue organization(Niessen, Leckband et al. 2011) and mechanosensing(Orré, Rossier et al. 2019), as well as cell migration and motility(Niessen, Leckband et al. 2011, Orré, Rossier et al. 2019), proliferation and

differentiation(Buckley, Rainger et al. 1998, Orré, Rossier et al. 2019). In many cases, molecular interactions depend not only on ligand-receptor affinity but also complexities that arise from the effects of avidity and multiple binding molecules attaching two surfaces. Further, the adhesive interactions are force dependent, both at the single molecule level and when multiple binding sites are in play. The classic case involves the attachment of leukocytes to endothelial cells through catch bonds whose binding strength can increase under load(Marshall, Long et al. 2003). In the respiratory tract, which motivates the development of our instrumentation system, adhesive interactions are essential to innate immune function, including pathogen trapping(Zanin, Baviskar et al. 2016, Hansson 2019). Mucus must have sufficent bind capacity to trap inhaled pathogens while simultaneously facilitating interactions between mucus and cilia that allow pathogen removal via mucociliary clearance (MCC). Adhesion between mucus and the underlying periciliary layer is implicated in MCC failure in obstructive pulmonary diseases(Button, Goodell et al. 2018, Hill, Button et al. 2022) but may also be critical for effective clearance against gravity(Carpenter, Lynch et al. 2018).

To investigate how molecular interactions contribute to MCC, we developed a micromagnetic bead-based adhesion assay applicable to cell culture and tissue explant assays, while also widely applicable to many bioadhesive systems. The system is high-resolution, scalable, and capable of capturing both equilibrium binding and mechanical detachment, enabling quantitative assessment of adhesive strength under biologically relevant conditions. The assay operates by exposing magnetic, ligand-coated beads to functionalized substrates and applying an upward magnetic force until detachment occurs. Because the force-dependent detachment events are captured dynamically via high-frame-rate microscopy, the assay enables calculation of the detachment force on a per-bead basis using Stokes drag as a real-time calibration. By varying bead and substrate chemistries and environmental conditions, this platform allows fine-grained assessment of adhesion strength response to both biochemical and rheological perturbations.

Prior bioadhesion assays include those that quantify the number of bound particles and those that quantify a force of attachment(Mackie, Goycoolea et al. 2017, Bayer 2022). We focus on force-based measurements to gain molecular insight into adhesive interactions and their modulation. The force measuring techniques range from macroscale systems that average over the millimeter length scale (or larger), and the single molecule detail available from atomic force microscopy(Sumarokova, Iturri et al. 2018). We endeavored to create an assay that resided in the middle ground, was amenable to high throughput implementation, and could be employed in the hydrodynamic environments of living systems such as MCC or blood flow.  Our system is compatible with live cell cultures and multiplexed well arrays, enabling throughput that rivals molecular assays while maintaining the physical relevance of detachment forces. Furthermore, the ability to map force distributions and extract binding mode subpopulations mirrors approaches used in force spectroscopy (Nathwani, Shih et al. 2018), providing a path toward quantitative interpretation of adhesion heterogeneity.

We demonstrate that our assay yields biologically meaningful measurements that can bridge biophysics and cell physiology. Within the context of mucoadhesion, such tools are critical not only for understanding mucus transport in health and disease but also for designing the next generation of mucoadhesive therapies, where tissue-specific retention, penetration, and clearance must be precisely engineered.

# Materials & Methods

## Multi-well Slide Fabrication

Custom 15-well slides were fabricated by combining vapor-phase silanization with a laser-cut adhesive well array. Standard glass slides were functionalized using 3-(triethoxysilyl)propylsuccinic anhydride (TEPSA), then hydrolyzed in water for 48 hours at 37 °C to homogenize the surface (Zhu, Lerum et al. 2012) and expose carboxyl groups for covalent ligand attachment. A 15-well mask, designed with fiducial markers for automated imaging, was cut from medical-grade adhesive using a laser cutter. Each well enclosed an 18 µL volume. The adhesive mask was affixed to the silanized slide and served to compartmentalize functionalization steps. Full fabrication parameters and silanization procedures are provided in Supplementary Material.

## Substrate Functionalization

Substrate functionalization was performed using standard carbodiimide chemistry to covalently link primary amines to activated carboxyl groups via NHS/EDC coupling. Slides functionalized with inert PEG or mucus derived from human bronchial epithelial (HBE) cultures (UNC Marsico Lung Institute) (Hill and Button 2012) were prepared by activating surface carboxylates on silanized, hydrolyzed slides.

Each functionalization and washing step used a 60 µL working volume, contained by the adhesive well walls and maintained by the surface tension of the solution. Activation was performed by incubating the slides in a solution of N-hydroxysuccinimide (NHS) and 1-ethyl-3-(3-dimethylaminopropyl)carbodiimide (EDC) in 50 mM MES buffer (pH 6) for 1 hour at room temperature on an orbital shaker (60 RPM). Activated wells were triple-rinsed with 50 mM MES before exposure to ligand (PEG or HBE mucus) at 1 mg/mL in PBS for 2 hours, also at room temperature on the shaker. Following binding, the wells were rinsed three times in PBS and stored in PBS at room temperature for use within 24 hours. Substrate glycoprotein content was validated using a modified Pro-Q™ Emerald 488 staining protocol, optimized for well-based formats and fluorescence imaging. Full staining procedures are provided in Supplementary Material.

## Bead Functionalization and Characterization

Fluorescent-yellow, 24 µm diameter carboxyl-modified magnetic beads (Spherotech, FCM-20052-2) were used in all adhesion experiments. Ligands were covalently coupled to the bead surface using standard NHS/EDC chemistry. Beads were functionalized with 2 kDa methoxy-PEG (mPEG; Creative PEGworks, PLS-269) or rhodamine-labeled PEG (RhoPEG; Creative PEGworks, PSB-2265) as negative controls, or HBE mucus as the experimental condition.

Beads were incubated at room temperature for 2 hours while rotating end-over-end with the selected ligand. To quench unreacted carboxyl groups, glycine was added, and the suspension was rotated for at least 30 minutes at room temperature. Beads were then washed three times by centrifugation at 6000 RPM for 15 seconds, decanting the supernatant, and resuspending in PBS.

A similar ligation was performed using rhodamine-labeled PEG (Sigma, 95172-250G-F) and imaged for successful binding (data not shown). For HBE mucus-coated beads, glycoprotein presence was validated using similarly-sized (20 µm diameter), nonfluorescent beads (Spherotech, CM-200-10) and the Pro-Q™ Emerald 488 Glycoprotein Gel and Blot Stain Kit (Thermo Fisher, P21875). A modified protocol optimized for bead suspensions was followed, where beads were fixed, oxidized, stained, and imaged under FITC conditions; full protocol details are provided in Supplementary Material. Zeta potential measurements were also conducted to confirm bead surface-modification using a density-matched medium to prevent bead sedimentation (see Supplementary Material).

## Sample Slide Preparation and Incubation

Each well of the functionalized substrate was filled with 60 µL of magnetic bead suspension, diluted 1:100 in transit medium, selected for its inert, non-interacting properties and a Newtonian viscosity suitable for Stokes force calibration. Here, a 10% w/w PEG (20 kDa) solution in PBS was used as the transit medium. The full rheological characterization is detailed in Supplementary Material.

The slide was incubated at room temperature for 2 hours to allow functionalized, magnetic beads to settle and bind to the immobilized analyte on the substrate surface. Following incubation, the sample slide was mounted on the microscope for data acquisition.

## Magnetics System

A custom "pincer" magnet geometry was constructed using two axially-aligned neodymium magnets (K&J Magnetics, B448-N52), each a rectangular prism (0.25 × 0.25 × 0.5 in) with a surface field of 0.53 T (Figure 1A & B). To focus the magnetic flux, tapered soft-iron tips were machined (UNC Department of Physics & Astronomy Instrument Shop) and mounted within a 3D-printed rig, producing a north-south pole gap of 200 µm (Lin and Valentine 2012). Hall probe measurements indicated a peak field strength of approximately 0.67 T within the pincer gap. This design has been shown to provide significant upward forces while also leaving the field of view clear of previously detached particles. Cone-shaped magnetic poles have the disadvantage of significant fluorescence background due to the presence of detached beads accumulating on the tip as it approaches the surface.

For controlled actuation, the magnet assembly (Figure 1C) was mounted on a kinematic breakaway mount (Thorlabs, model BK-2A) and translated along the optical axis via a stepper motor (Thorlabs, Z25b). The magnet was advanced toward the substrate at 0.2 mm/s, starting 12 mm above the surface (Figure 1D). The z-position zero point was determined by using a second sample slide, dry-coated with 1 µm red-fluorescent beads (Thermo Fisher, F13083) and lowering the magnetic assembly until visually observing the deflection of the glass surface.

Each video included a 60 s data collection interval with an additional 5 s dwell once the magnet contacted the substrate. At the maximum distance (12 mm), no detachment was observed for COOH-modified beads on a COOH-functionalized substrate, confirming that initial force levels were sub-threshold. Lateral crosstalk from stray fields was negligible at the tested well spacing and magnet dimensions.

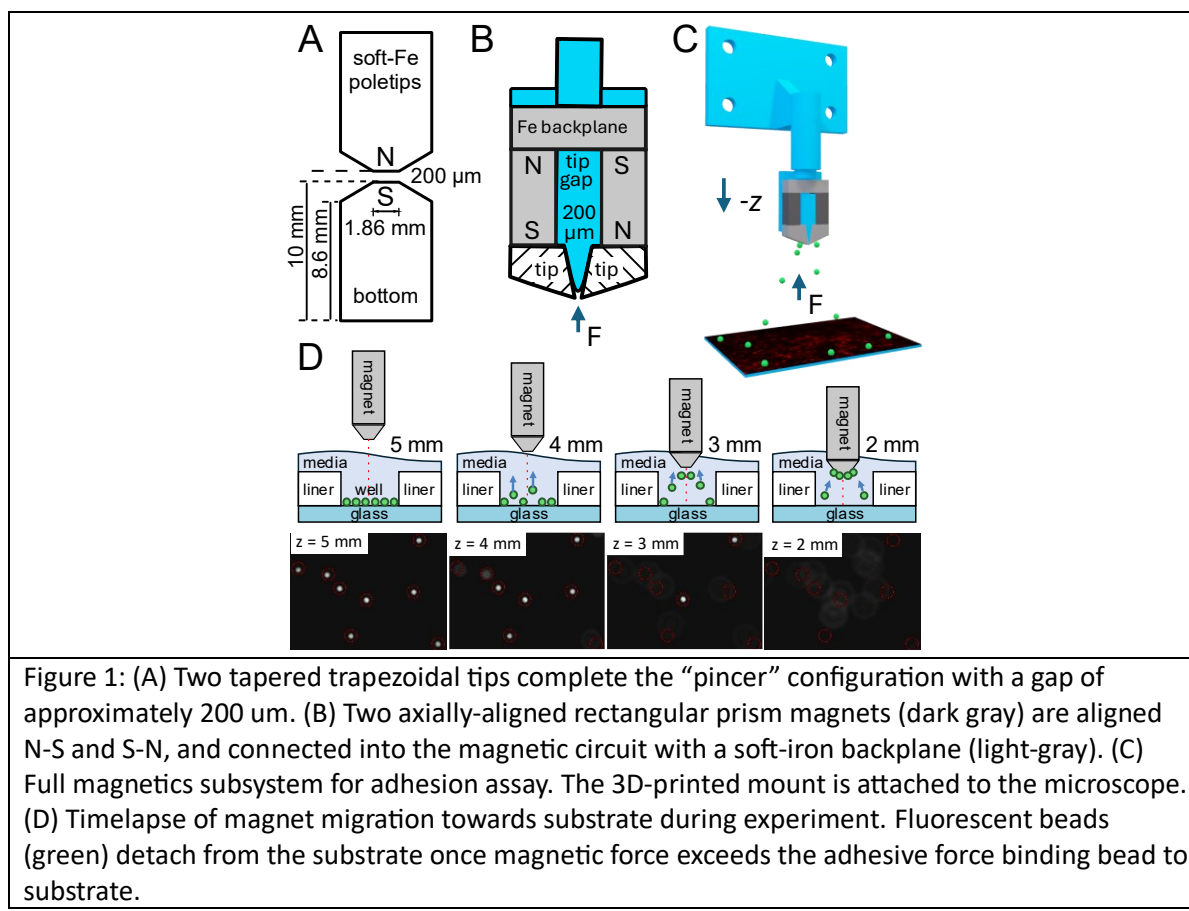

Figure 1: (A) Two tapered trapezoidal tips complete the "pincer" configuration with a gap of approximately 200 um. (B) Two axially-aligned rectangular prism magnets (dark gray) are aligned N-S and S-N, and connected into the magnetic circuit with a soft-iron backplane (light-gray). (C) Full magnetics subsystem for adhesion assay. The 3D-printed mount is attached to the microscope. (D) Timelapse of magnet migration towards substrate during experiment. Fluorescent beads (green) detach from the substrate once magnetic force exceeds the adhesive force binding bead to substrate.

## Microscopy

Bead imaging was performed using a Nikon TE-2000 inverted optical microscope equipped with epifluorescence filters for rhodamine (excitation/emission: 544 nm/576 nm) and FITC (490 nm/525 nm). Fiducial alignment markers on the sample tray were imaged using a 4x objective (Nikon MRH00045, NA = 0.13), while bead tracking was conducted using a 10x objective (Nikon MRH00105, NA = 0.30). Calibration using a stage micrometer yielded pixel scaling factors of 1.72 µm/pixel for 4x and 0.692 µm/pixel for 10x imaging.

Software for controlling the microscope illumination and focus, XY stage translation, magnet position/distance, and image acquisition was written in MATLAB (available upon request). Where appropriate, software packages containing hardware control sequences were acquired through MatlabCentral and modified to function within the adhesion assay workflow.

An automated, closed-loop XY translation stage (Ludl MAC6000) was used to position the microscope field of view over each well in the sample array. Well center positions were calculated by linear interpolation from the fiducial marks based on the original design file. A calibration tray—containing fiducial-sized cutouts at each intended well center—was used to validate coordinate transformation accuracy. Across 10 repeat visits to each well location, the positional deviation remained below 25 µm, with negligible hysteresis or cumulative error.

Bead motion was recorded using a FLIR/Point Grey Grasshopper3 camera (GS3-U3-32S4M-C) operating at a resolution of 1024 × 768 pixels and 16-bit depth. Illumination intensity was adjusted to place peak pixel values near 90% of the dynamic range for an 8 ms exposure time. Images were acquired at 125 frames per second over a 60-second interval to capture bead trajectories during magnetic actuation.

When necessary to image the full surface area of each well, the XY stage was programmed to traverse the calibrated coordinate grid while collecting mosaic-style tiled images using the 10× objective. Individual tiles were reassembled into composite images via software stitching. These mosaics enabled approximate bead counts before and after magnetic actuation. Full-slide mosaics were acquired immediately prior to and following video collection.

### Bead Tracking and Z-Displacement Calibration

To track bead motion in the z-direction (normal to the substrate), we calibrated bead image morphology as a function of defocus. A 2% stock bead solution was diluted 1:100 in anhydrous ethanol and dried onto a 24 × 40 mm glass coverslip. The coverslip was mounted on a motorized z-stage via a kinematic breakaway mount and submerged in transit medium to replicate experimental imaging conditions. Beads were imaged at 1 μm intervals as they were stepped away from the focal plane defined at the substrate surface.

At each step, bead images were mapped radially from their centers under the assumption of rotational symmetry, effectively "unwrapping" bead pixels to build a reference library of defocus-based point spread functions (PSFs). This PSF mapping enabled precise z-displacement tracking during detachment. Cross-validation using separate calibration beads yielded z-position errors of less than 1 μm on average.

Bead positions were tracked in three dimensions using Video Spot Tracker software (https://github.com/CISMM/video). Z-displacements were mapped using the empirically derived defocus-based point spread function, calibrated at 1 μm resolution. Only the linear portion of each trajectory immediately following detachment was retained. The slope of this segment yields the bead velocity, and fitting error from the linear regression was used to weight the reliability of each force estimate.

### Quantitative Analysis Workflow

The analytical workflow for the adhesion assay is shown as a flow diagram in Figure 2. Once the spread function is used to calibrate bead position as a function of defocus, bead trajectories are tracked during detachment using high-speed microscopy. Bead velocity post-detachment is used to calculate force via Stokes' law on a per-bead basis. Cumulative force distributions are constructed and fitted using complementary error functions to resolve distinct adhesive modes. Bootstrapping provides confidence intervals, and weak or low-abundance modes are excluded from analysis.

### Force Calculation from Bead Velocity

There are two clear factors that can hinder a calibration procedure that relies on force calibration prior to the experiment. First, we seek to acquire data from a field of view that is large with respect to the force gradients in the system, which means that forces on the beads are heterogenous across the microscope field of view. Second, there is variation in the bead-to-bead magnetic content. We therefore set out to establish a calibration procedure that operates on a bead-by-bead basis within the experiment. For each bead, the applied force at the moment of detachment was calculated using Stokes' law, $F = 6\pi\eta a v$, where $a$ is the bead radius, $\eta$ is the medium viscosity, and $v$ is the measured post-detachment velocity. Given constant bead size and known medium viscosity, magnetic material content per bead becomes irrelevant, as each bead's velocity serves as an internal force calibration. For our system, the maximum measurable z-displacement during bead detachment is 250 µm, corresponding to a peak velocity of 31 mm/s. Given the measured transit medium viscosity of 12 mPa·s (Supplementary Material), this results in a maximum resolvable force of 85 nN using Stokes' law.

### Curve Construction and Mode Fitting

To interpret detachment force data, we assume that each adhesive interaction produces a Gaussian-distributed range of pull-off forces across the bead population. Since we measure the cumulative number of beads remaining on the substrate as force increases, the resulting curves correspond to integrated force distributions mathematically represented as complementary error functions. This approach mirrors the analytical framework used in earlier adhesion assays by Lauffenburger and Hammer, where population-level detachment events were modeled using integrated probability distributions to resolve underlying binding modes(Kuo and Lauffenburger 1993, Kuo, Hammer et al. 1997).

We pool bead forces from three replicate wells per condition, sort them in ascending order, and then assemble as a cumulative force distribution ("force curve") for each bead-substrate pairing. The distribution was modeled as the sum of up to two complementary error functions, where each mode is defined by scale factors, $a_i$, center forces, $b_i$, and spread factors, $c_i$. The total mode scales were constrained to a sum of one. The fitting equation becomes:

$$f(F_d) = \sum_{i=1}^{N} \frac{a_i \operatorname{erfc}\left(\frac{\sqrt{2}(F_d - b_i)}{2c_i}\right)}{2}$$

This approach enables resolution of heterogeneous interaction populations, such as subpopulations of weak or incomplete binding events. When more than one mode is present, they are considered separable only if the sum of their spreads is less than the distance between their respective peak forces.

Fitting was performed using a genetic algorithm (MATLAB Global Optimization Toolbox), which required no initial parameter estimates and reliably converged to a global minimum. While computationally slow, this method provided robust fits to complex force distributions.

### Bootstrapping Confidence Intervals

To estimate uncertainty in fitted parameters, a bootstrapping procedure was applied to each force curve. Resulting parameter distributions were used to compute 95% confidence intervals. Modes contributing less than 20% of the total signal were excluded from downstream analysis.

Final force values were summarized using standard descriptive statistics. Group comparisons were assessed using ANOVA with Bonferroni correction for multiple comparisons. All reported adhesion data were derived from at least three independent biological replicates, each representing separate HBE mucus preparations on separate substrates.

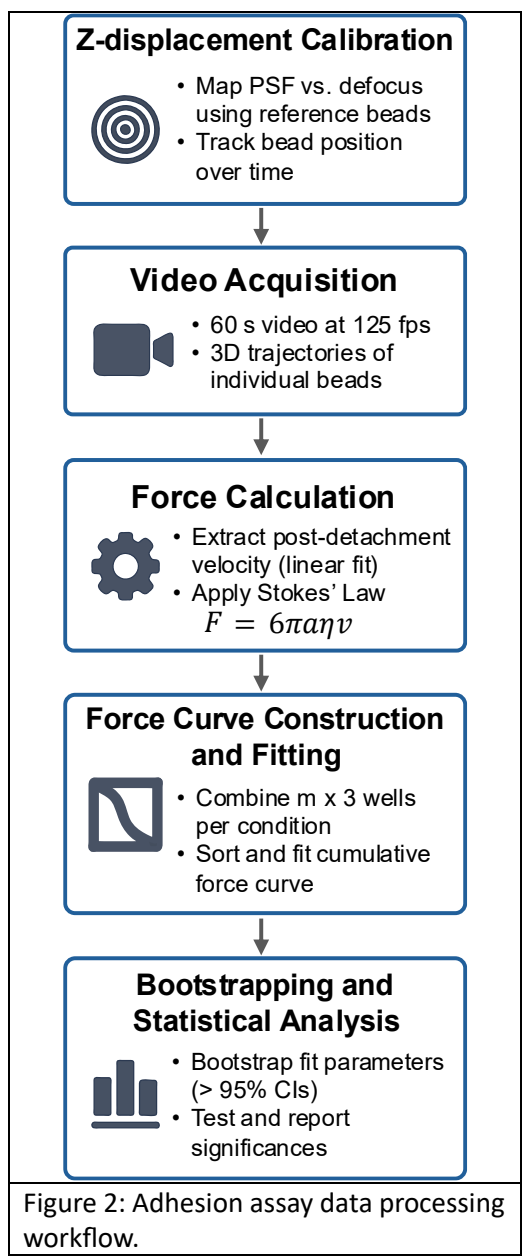

Figure 2: Adhesion assay data processing workflow.

## Results

### Characterization of Bead Functionalization

To confirm successful bead surface modification with HBE mucus, we stained HBE-coated and glycine-coated non-fluorescent control beads with Pro-Q™ Emerald 488 glycoprotein stain and imaged large numbers of them as mosaics under identical exposure and scaling conditions. As shown in Figure 3A, sample bead images revealed stronger fluorescent signal in HBE-coated beads (right) compared to glycine-coated beads (left), with peak pixel signal over corrected-background reaching 70 and 45, respectively, at the same scaling.

Both bead types exhibited a central signal depression ("caldera"), likely resulting from focus being set at the bead center and enhanced edge brightness caused by the higher projected fluorophore density at regions where the bead surface intersects the optical axis at a steep angle. To account for this, the caldera region was removed from analysis with a mask, and the surrounding fluorescent "ring" was used to calculate corrected average intensities for individual beads (Figure 3B). The resulting distributions (Figure 3C) show a clear shift toward higher fluorescence signal in the HBE-coated population, confirming the presence of glycosylated mucins on the bead surface.

Surface modification was further validated by zeta potential measurements (Figure 3D). Bare COOH-functionalized beads showed the highest net negative charge. Coating with PEG or HBE mucus reduced the magnitude of the negative surface charge, consistent with masking of carboxyl groups and successful ligand attachment in each case.

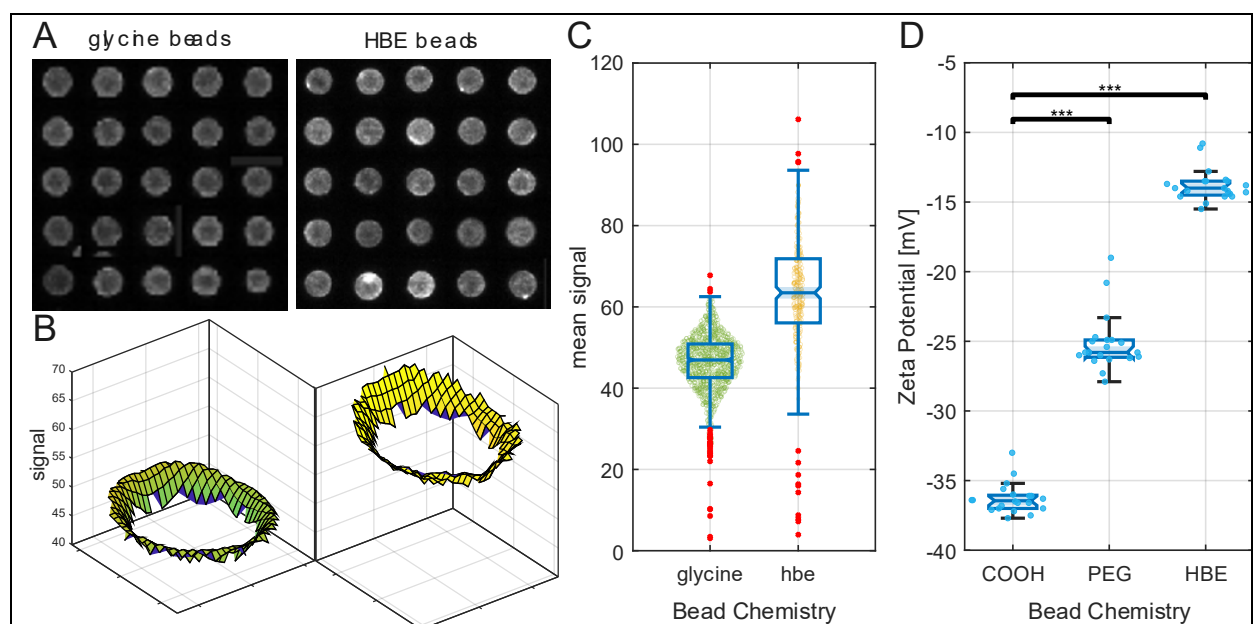

Figure 3: Validation of bead surface functionalization. (A) Sample fluorescent images of glycine-coated (left) and HBE-coated (right) beads stained with Pro-Q™ Emerald 488. Images are displayed with identical intensity scaling and z-axis location. (B) Corrected bead signal maps for glycine (left) and HBE mucus (right) after excluding central "caldera" regions; ring intensities were used to generate bead-specific average signals. (C) Distributions of mean fluorescent signal for all glycine- (N= 864) and HBE-coated (N=302) beads, showing higher corrected intensity in the HBE-coated population. (D) Zeta potential measurements of all bead types used in the study. COOH-functionalized beads exhibit the most negative charge, while PEG and HBE-coated beads show progressively reduced surface charge, consistent with successful surface modification.

## Substrate Functionalization

Cropped mosaic images collected from all 15 wells are shown in Figure 4A, with entire mosaics shown in Supplementary Material. The upper-left position, well 1, received neither mucus nor stain, and served as a background reference. Wells 2 through 5 were stained with Pro-Q™ Emerald 488 but lacked HBE mucus, allowing quantification of unbound dye fluorescence. These wells exhibited low signal overall but were marked by small, bright puncta—likely due to precipitated or crystallized dye. Notably, well 2 showed elevated signal in its lower half, resulting from unintended HBE mucus contamination by adjacent well 7 during the coupling step.

Wells 6 and 11 contained HBE mucus but no stain. These exhibited no elevation in signal intensity relative to the unstained background (well 1), indicating minimal intrinsic mucus autofluorescence. The remaining wells (7-10 and 12-15) were treated with both HBE mucus and Pro-Q™ Emerald 488. Among them, well 9 contained a horizontal bright streak consistent with fiber contamination.

To quantify signal intensity across wells, pixel-wise intensity histograms were generated and are shown in Supplementary Material, where median pixel values for each well are indicated by a star glyph. Wells with stain alone (no mucus) showed rightward shifts in histogram peaks relative to the background, indicating weak fluorescence of unbound dye. Wells with both stain and mucus shifted further rightward, consistent with successful staining of surface-bound mucins. Peak widths varied across wells, potentially reflecting heterogeneity in mucus distribution or surface coverage.

Boxplots of median signal intensity values, overlaid with individual well-level measurements as swarm plots (Figure 4B), further support successful and spatially consistent coating of the substrate surface with HBE mucus.

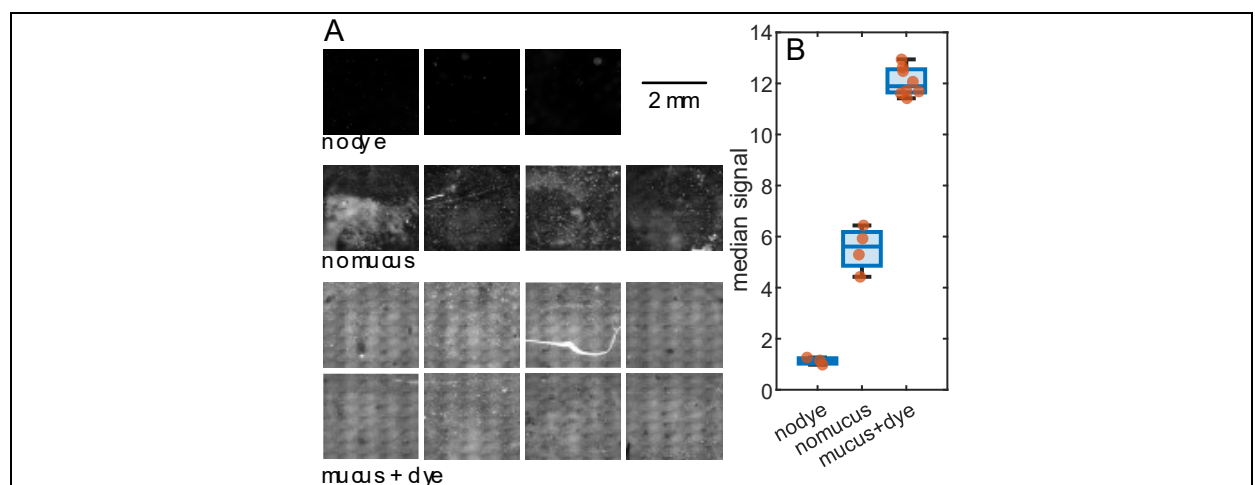

Figure 4. (A) Whole-slide validation of HBE mucus coating using Pro-Q™ Emerald 488 staining. Each image comprises a cropped mosaic composite across the full 7 mm well. Full slide imaging is available in Supplementary Material. Cropped mosaic fluorescence images of all 15 wells from a representative slide. Row 1 (top) received unstained mucus and serves as background reference. Row 2 contains stain only, showing low-intensity signal with occasional bright puncta likely due to dye precipitates. Rows 3 and 4 were treated with both mucus and stain. (B) Boxplots of median pixel intensity per well, with overlaid swarm chart showing individual well-level values. Dye-only wells (nomucus) show modest signal elevation, while mucus-plus-dye wells exhibit significantly higher shifts in peak intensity, confirming consistent and detectable HBE mucus coating across treated wells.

## Z-tracking Calibration and Depth Resolution

To enable quantitative 3D tracking of bead motion, we performed a z-axis calibration of bead image morphology as a function of defocus. Figure 5A (left) shows an XZ cross-section constructed from a stack of bead images captured on-axis at 1 µm intervals, spanning from in-focus (0 µm) to a maximum defocus of 300 µm. As expected, the image intensity decreased, and spatial spread increased with defocus. Representative snapshots at selected depths (Figure 5A, right) highlight key stages of image degradation across the defocus range. At 250 µm—the empirical tracking limit for a 50-pixel tracking radius—the bead image remains detectable but highly diffuse.

Radial intensity profiles further quantify this evolution (Figure 5B). The central pixel intensity and that at 5 µm radius decrease monotonically with defocus, while intensity at 10 and 15 µm radii increases before falling off at greater depths, consistent with broadening of the bead's point spread function. These intensity trends form the basis of our defocus-based depth tracking algorithm.

A calibration curve relating z-position to axial displacement is shown in Figure 5C. A brief tracking sensitivity below ~15 µm arises when the focal plane intersects the bead. In practice, this region is avoided by initializing the focal plane at the bead-substrate contact point. This calibration supports accurate z-tracking over a range of up to 250 µm with 1 µm precision.

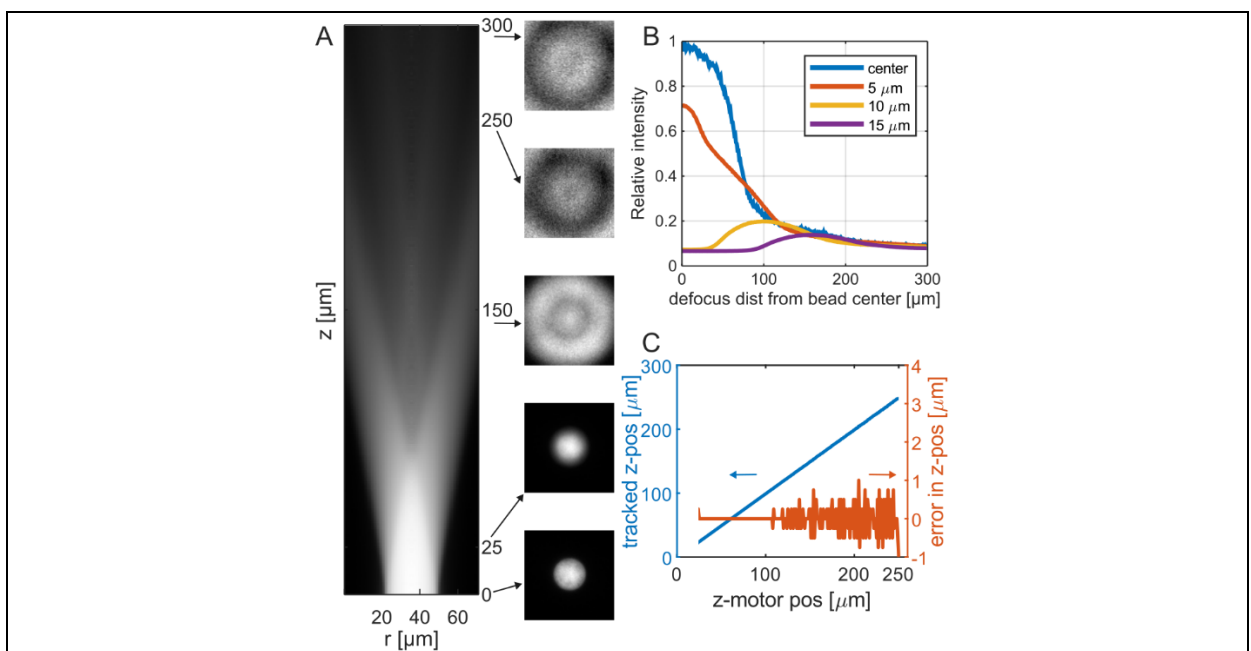

Figure 5: (A) Left: XZ spread image function of a 24 µm fluorescent bead during axial defocus calibration, with depth increasing from bottom (in-focus) to top (300 µm defocus). As defocus increases, signal intensity decreases and spatial spread increases. Right: Representative bead images at selected defocus depths: 25 µm (~bead diameter), 150 µm (mid-range), 250 µm (tracking limit for 50-pixel radius), and 300 µm (maximum measured defocus). Image intensities are scaled for visibility. (B) Intensity profiles sampled at the bead center and radial distances of 5, 10, and 15 µm. Central intensity decreases monotonically with defocus, while peripheral intensity increases then decreases, reflecting the evolving point spread function shape. (C) Z-position calibration curve and tracking error relating defocus-derived bead position to physical z-displacement.

## Detachment Forces

To validate the adhesion assay, we tested control slides with bead-substrate functionalization in all wells for three such conditions: COOH-COOH, PEG-PEG, and HBE-HBE, using 10% 20 kDa PEG in PBS as a non-interfering transit medium. Figure 6A illustrates the multi-mode error function used for fitting force curves, where each data point presented in Figure 6B and 6C represents a bead's single detachment force The detachment profile for COOH-COOH curves reveals two distinct force modes in the interaction, where each mode is responsible for approximately half of the bead count. Both modes are reproducible and separable across independent measures, labeled as COOH-Low (68 pN) and COOH-High (660 pN) in Figure 6B. Once separated, dispersion in force for the different conditions becomes more consistent.

All four interaction types, along with their reproducibility, are plotted in Figure 6C. The COOH-COOH-Low mode (68 pN) and the predominantly single-transition PEG-PEG interaction (1.6 nN) differ by more than an order of magnitude in force. The remaining two curves, representing independent HBE-HBE measurements, demonstrate the strongest detachment forces at 6 nN with tight reproducibility.

Figure 6D summarizes the detachment force curves using a fitting equation that accommodates up to two interaction modes. Since the two COOH-COOH modes are both separable and statistically distinct, we treat them as independent datasets. This approach reveals statistically significant differences ($p < 0.05$, Bonferroni-corrected) among all control conditions: COOH-COOH, PEG-PEG, and HBE-HBE. The two HBE-HBE slides were statistically indistinguishable ($p > 0.05$), supporting the platform's reproducibility and functionality across bead and substrate preparations.

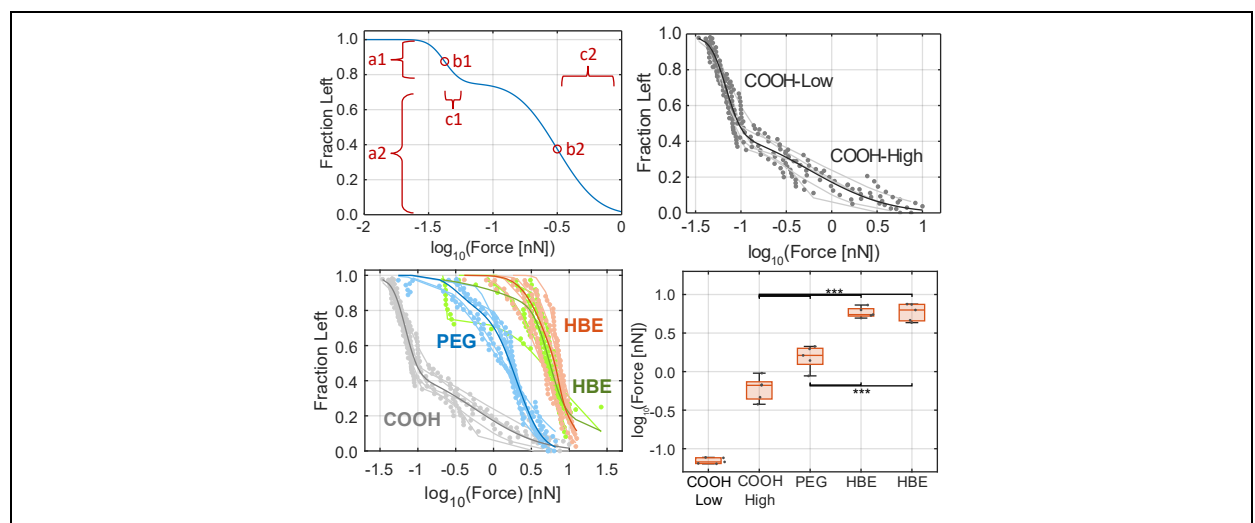

Figure 6: (A) Fitting equation used analyze detachment force curves in the adhesion assay. (B) Cumulative force curves for COOH–COOH interactions reveal two distinct adhesive modes, COOH–Low at 68 pN and COOH-high at 660 pN. (C) Reproducible force curves across four interaction types: COOH–COOH, PEG–PEG, and HBE–HBE. (D) Summary of fitted mode forces with statistical comparisons; all conditions differ significantly ($p < 0.05$, Bonferroni), except for the two HBE–HBE replicates. Significance bars are shown for only a highlighted subset of comparisons.

## Discussion

Adhesion phenomena between an approaching particle and the substrate can be mediated by several interactions, including electrostatic, dispersion, polymer entanglement, and ligand-receptor binding. Our measurements of bead-substrate interactions show that the COOH-COOH pairing yields the lowest threshold force, which is consistent with electrostatic repulsion of similarly charged surfaces since both bead and substrate should have net negative charge at buffer pH. The PEG-PEG interaction marks our next strongest, which supports literature that considers PEG polymers as weakly interacting. Finally, our strongest interaction is between our mucus (HBE) coated beads and similarly coated substrates. While this interaction can be complex, we next place this measurement into the context of receptor-ligand binding that may be mediated by lectins present in the mucus that recognize glycans on the heavily glycosylated mucin chains. These lectins are often multivalent and may act as cross-linkers between mucin chains on the bead and substrate.

The adhesion assay permits quantitative interpretation of the measured detachment force in terms of receptor and ligand concentrations and the free energy of interaction or dissociation constant ($K_D$), a parameter often obtained by techniques such as surface plasmon resonance (SPR). Bead detachment occurs when the applied magnetic force acting over a distance exceeds the adhesive free energy holding the bead to the substrate. This work of adhesion, $W$, can be approximated as the product of force and bond extension distance:

$$W = Fa$$

where $F$ is the measured force and $a$ is the molecular-scale bond extension distance.

Following the theoretical framework by Kuo and Lauffenburger, the relationship between detachment force and dissociation constant is given by:

$$F_t = R_c \frac{2k_B T}{a} \ln\left[1 + \frac{N_L}{\eta K_D}\right],$$

where $R_c$ is the receptor number in the bead contact region, $k_B T$ is the thermal energy scale (4.1 pN*nm), $a$ is the bond extension limit (~ 1nm) (Haugstad, Stokke et al. 2015, Hadjialirezaei, Picco et al. 2017), $N_L$ is the ligand density on the substrate, and $\eta$ is a conversion factor to relate 2D to 3D dissociation constant (Kuo and Lauffenburger 1993).

Our measurement of a threshold detachment force of about 5 nN for the interaction of HBE-coated, 24 µm diameter beads with a flat substrate coated with HBE mucins can be placed into this framework. In the simplest sense, we can approximate the receptor number with estimates for contact area and ligand density. If we assume a contact distance of 50 nm, the spherical cap area "in contact" with the substrate, we calculate an area of $A = 2\pi(12\mu m)(0.05\mu m) = 4\mu m^2$. If we assume a receptor area of ~1000 nm² per receptor, we calculate a receptor number (on the bead) in the contact region of $Rc \approx 4000$. We can use our estimate for the receptor area to assume a similar density for the ligands on the substrate and estimate $N_L \sim 10^3 \mu m^{-2}$. The term $\eta$ is a conversion factor to relate 2D to 3D dissociation constant and has been estimated to be $\eta \sim 1.3 \times 10^{18} liter/cm^2 mol$. Adhesion between HBE-coated beads and bead and substrates involves multiple potential interactions, including the presence of lectins that have been detected within mucus pools similar to those from which we derive our mucus to coat our surfaces. The $K_D$ for putative lectins can vary over several orders of magnitude, but since $K_D$ appears within the argument of a logarithmic function, the resulting detachment force is less sensitive to this quantity. To continue our estimate, we assume a $K_D \sim 5 \times 10^{-12} M$, we obtain $F_t \sim 5 nN$. We also note that our estimates are consistent with a single molecule threshold detachment force that is in the range of several pN, in the range of single molecule force experiments (Haugstad, Stokke et al. 2015, Hadjialirezaei, Picco et al. 2017).

This calculation provides context for situating our assay within broader models of multivalent receptor-ligand binding in adhesion. Similar computations show that for constant ligand/receptor densities, the threshold force can vary between 80 nN and 5 nN by varying Kd between $10^{-15}$ M and $10^{-10}$ M, a range that covers strong to weak lectin binding interactions to sialic acid moieties on mucins. It is interesting to note that the threshold force is sensitive to the receptor number, and therefore interference assays that interrupt binding through the interaction of competing species may be useful in elucidating adhesive interactions using our method.

Comparing these initial data to previous studies of mucus adhesion and molecular interactions, we note that our data offer a complementary approach to the existing measurements of Button et al. (Button, Goodell et al. 2018) that explore the role of mucus hyperconcentration on mucus adhesion and cohesion. By controlling the surface chemistry of our magnetic particles, we can probe the strength of specific chemical interactions that may contribute to mucus adhering to the underlying epithelial layer, making the system well suited for studies involving small molecule or lipid nanoparticle formulations intended to penetrate the mucus barrier for therapeutic delivery. It is equally applicable to the design of mucoadhesive therapeutics aimed at increasing residence time in the lungs or nasal passages (Chaturvedi, Kumar et al. 2011). Likewise, by tuning the density of surface-bound molecules on probe particles, our system can support studies into how mucosal layers trap pathogens and inhaled particulates through the collective effect of many low-affinity binding interactions (Wessler, Chen et al. 2016).

## Conclusion

We have developed a high-resolution, scalable adhesion assay capable of quantifying molecular detachment forces under biologically relevant conditions. By coupling micromagnetic actuation with precision imaging and customizable surface chemistry, this platform enables detailed measurement of adhesion strength across a range of biointerfaces. Validation against inert controls and mucus-coated surfaces confirmed the assay's sensitivity and reproducibility. The ability to resolve force modes and extract binding heterogeneity positions this system as a useful tool for studying multivalent interactions in complex biological systems. Future work will extend this framework to explore specific molecular interactions, including lectin-glycan binding, and assess the effects of chemical interference by modulating the composition of the bead transit medium. These capabilities will support a broad spectrum of applications, from mucosal immunology to drug delivery.

## Acknowledgments

This work was supported by the Cystic Fibrosis Foundation under grant numbers SUPERf2020G0, HILL20Y2-OUT, and SUPERf2022G0. The authors acknowledge the Mehmet Kesimer lab for our use of their zeta sizer. We thank Megan Pramojaney for contributions made to instrumentation control software and automation routines used in the adhesion assay platform. We also acknowledge Autumn Samson for efforts made in early method development during the initial stages of the project.

## Author Declarations

The authors declare that they have no conflicts of interest. No financial, personal, or professional relationships influenced the design, execution, or reporting of this work.

## Data Availability

The data that support the findings of this study are available from the corresponding author upon reasonable request.

## Supplementary Material

### Substrate Silanization

Standard 50 × 75 × 1 mm glass slides served as substrates and were first cleaned in an oxygen plasma chamber for 15 minutes. Cleaned slides were immediately placed in the center of a 150 mm diameter Petri dish containing four 13 mm aluminum weighing pans positioned at the corners. Each pan was loaded with 60 µL of a 1% v/v solution of 3-(triethoxysilyl)propylsuccinic anhydride (TEPSA) in toluene. The Petri dish was sealed and incubated overnight at 80 °C to promote vapor-phase silanization.

Following silanization, slides were hydrolyzed in deionized water at 37 °C for 48 hours to convert succinimide rings to carboxylate groups, yielding two –COOH moieties per silane (Zhu, Lerum et al. 2012). Contact angle measurements, performed using in-house instrumentation (data not shown), confirmed successful surface functionalization in early validation experiments (Behroozi and Behroozi 2019).

### Multi-Well Slide Fabrication

A 15-well adhesive template was designed in Adobe Illustrator and fabricated using a laser cutter (Universal Laser Systems, VLS6.6) located in the UNC BeAM makerspace. Laser power and speed were set to 75% and 100%, respectively. Fiducial markers (500 µm diameter) were cut at each corner of the array to enable XY calibration of the slide position during automated imaging. The fiducial center-to-center spacing was 69.5 mm (X) and 44.5 mm (Y), while well center spacing was 11.8 mm (X) and 16.2 mm (Y). Well center positions were calculated by linear interpolation between fiducial marks, matched to the design coordinates used during laser cutting.

Each well accommodates a sealed volume of 18 µL. Prior to assembly, silanized and hydrolyzed glass slides were rinsed with isopropyl alcohol and dried. The laser-cut adhesive layer (Adhesives Research, 90106NB) was then aligned and affixed to the slide surface. All subsequent substrate modifications were performed within the defined wells of this array.

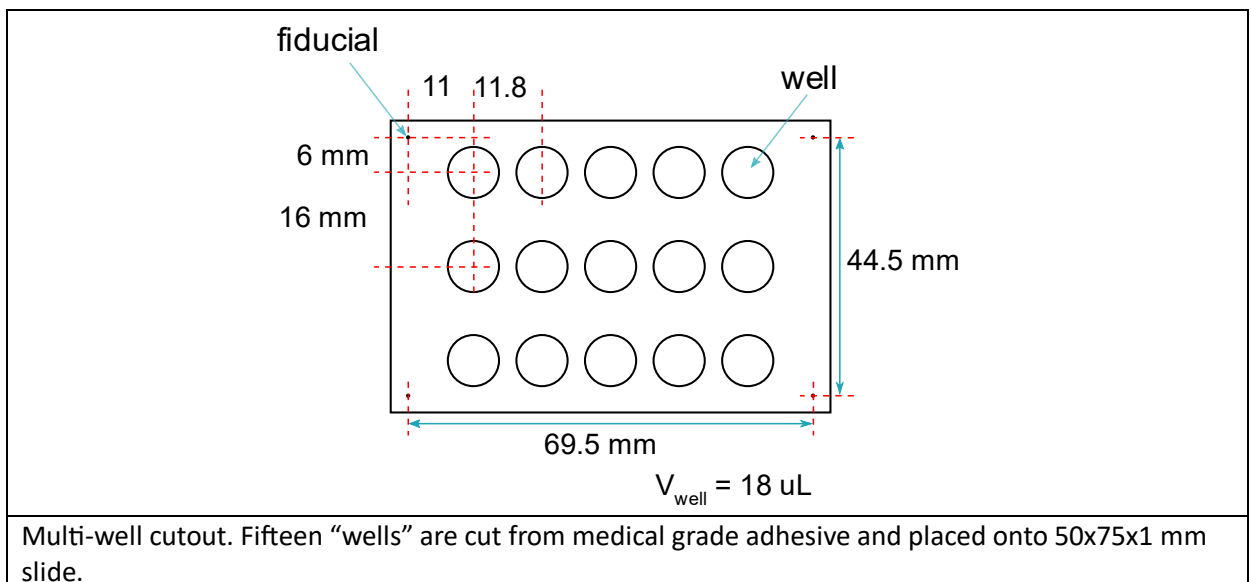

Multi-well cutout. Fifteen "wells" are cut from medical grade adhesive and placed onto 50x75x1 mm slide.

## Labeling HBE-functionalized Substrate

To confirm successful immobilization of HBE-derived mucus on the slide substrate, we performed fluorescent staining using the Pro-Q™ Emerald 488 glycoprotein stain (Thermo Fisher, P21875), adapted from the manufacturer's protocol for use in adhesive well arrays. Slides were fixed in 50% methanol with 5% acetic acid, oxidized using periodic acid in 3% acetic acid, and incubated with a filtered Pro-Q™ 488 staining solution prepared in DMSO. All incubation steps were performed on a low-speed orbital shaker, with the slide kept in a dark enclosure from the staining step onward to prevent photobleaching. Stained slides were washed with deionized water, then methanol, air-dried, and imaged using FITC filter settings. Care was taken throughout to preserve the integrity of the well surface and avoid artifacts from dye precipitation or photobleaching. This protocol provided reliable contrast between HBE-coated and control wells and was used in full-slide imaging experiments.

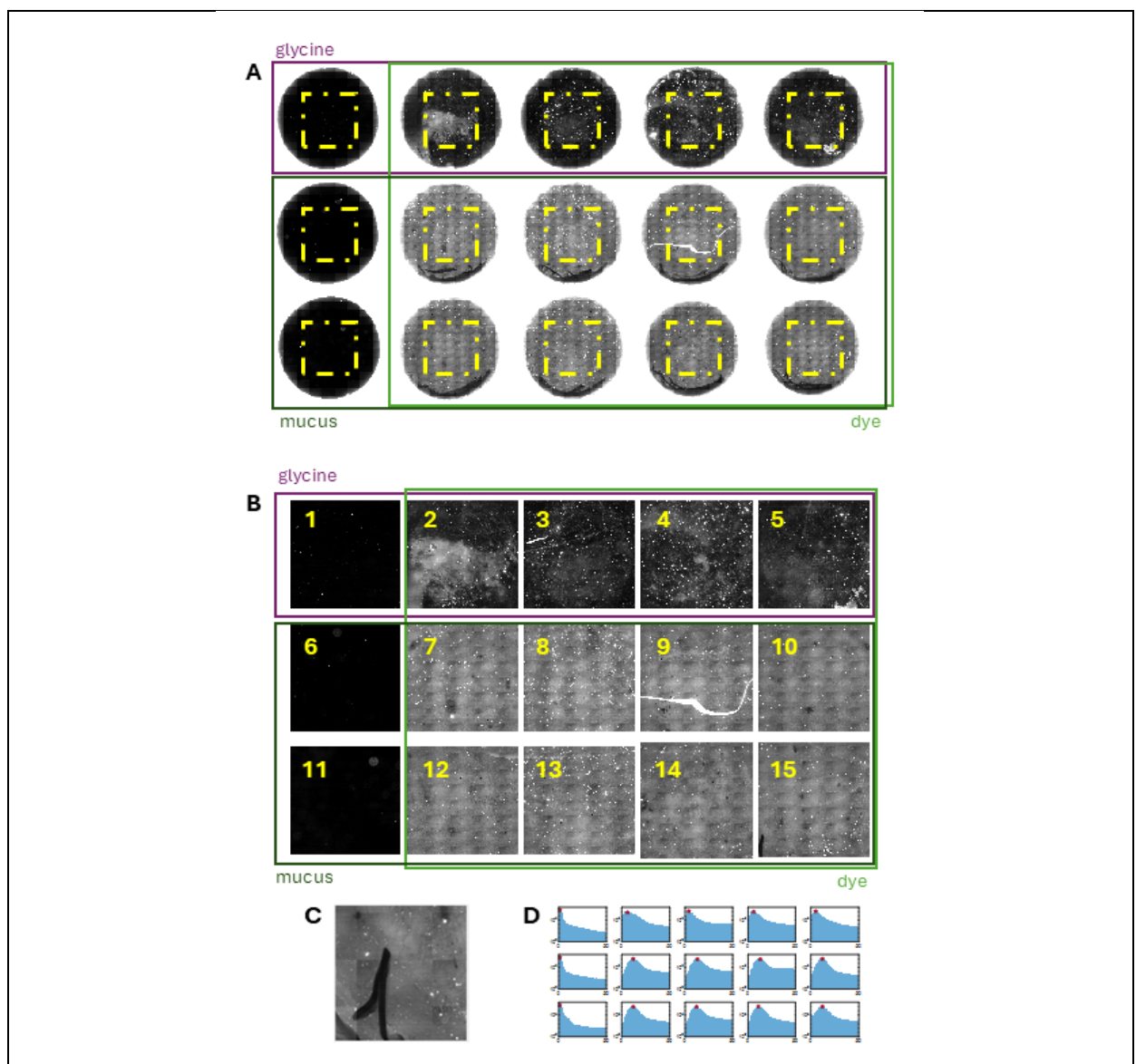

(A) Mosaic images for wells on TEPSA-treated glass substrate and various imaging conditions with HBE mucus and/or ProQ-488 dye. Cropping window marked by yellow square annotation. (B) Cropped mosaics for wells 1 – 15 (labeled in yellow). Well 1 has neither mucus nor dye. Wells 2 – 5 contain dye with no mucus. Wells 6 and 11 contain mucus with no dye. Wells 7 – 10 and 11-15 contain both HBE mucus and dye. Punctate areas of brightness occur due to aggregated or concentrated precipitates of the dye molecule. "Cloudy" region in Well 2 corresponds to mucus spillage into well during NHS and EDC ligation. Bright arc of expression in Well 9 is caused by errant fiber. (C) Intentional scrape with pipette tip through mucus-coated substrate in Well 15. (D) Pixel histograms for each well with the median value annotated by star glyph.

## Labeling HBE-functionalized Beads

To validate surface glycoprotein content on HBE-functionalized beads, we adapted the Pro-Q™ Emerald 488 glycoprotein stain (Thermo Fisher, P21875) protocol for bead suspensions. Non-fluorescent carboxylated beads (20 μm, Spherotech CM-200-10) were used to enable imaging in the FITC channel.

18Beads were washed and incubated with Fix Solution (50% methanol, 5% acetic acid) for 1 hour under gentle end-over-end mixing. After three washes in acetic acid buffer, beads were oxidized with periodic acid and then stained for 1 hour with filtered Pro-Q™ 488 stain diluted 1:10 in DMSO. All steps post-staining were carried out in light-protected conditions to minimize photobleaching. Beads were then washed in water and methanol, resuspended in a small volume of methanol, and drop-cast onto a clean glass slide. Rapid drying prevented dye diffusion artifacts, and bead fluorescence was imaged under standard FITC filter settings.

### Bead zeta potential

Zeta potential was measured using a Malvern Zetasizer Nano series instrument equipped with a folded capillary zeta cell. To prevent sedimentation of the high-density magnetic beads (1180 kg/m$^3$), measurements were performed in a density-matched medium consisting of 0.5 M sucrose dissolved in PBS, maintaining consistent ionic strength while stabilizing the suspension.

Beads were initially suspended in PBS and required exchange into the measurement medium. Samples were centrifuged at 6000 RPM for 15 seconds, the supernatant was removed, and the pellet was resuspended in 0.5 M sucrose-PBS solution. The resulting suspension was loaded into the folded capillary cell, and 20 consecutive measurements were acquired per sample. This procedure was repeated for each bead surface treatment to ensure consistency across conditions.

### Transit Medium Rheology

The bead translation medium was prepared by dissolving 20 kDa polyethylene glycol (PEG) at 10% w/w in 1x PBS. Rheological measurements were conducted using a stress-controlled cone-and-plate rheometer (TA Instruments AR-G2) equipped with a 40 mm/1° cone. A stepped flow protocol was performed at room temperature (23 °C), without pre-shear, over a stress range from 1 to 30 Pa. Following the flow test, a creep step was applied at 1 Pa for 5 minutes. No elasticity was observed in the tested range. Viscosity was determined from the slope of the creep curve and compared with stepped flow results, yielding a mean value of 12 ± 0.2 mPa·s.

## References

Bayer, I. S. (2022). "Recent Advances in Mucoadhesive Interface Materials, Mucoadhesion Characterization, and Technologies." Advanced Materials Interfaces **9**(18): 2200211.
Behroozi, F. and P. S. Behroozi (2019). "Reliable determination of contact angle from the height and volume of sessile drops." American Journal of Physics **87**(1): 28-32.
Buckley, C. D., G. E. Rainger, P. F. Bradfield, G. B. Nash and D. L. Simmons (1998). "Cell adhesion: More than just glue (Review)." Molecular Membrane Biology **15**(4): 167-176.
Button, B., H. P. Goodell, E. Atieh, Y. C. Chen, R. Williams, S. Shenoy, E. Lackey, N. T. Shenkute, L. H. Cai, R. G. Dennis, R. C. Boucher and M. Rubinstein (2018). "Roles of mucus adhesion and cohesion in cough clearance." Proc Natl Acad Sci U S A **115**(49): 12501-12506.
Carpenter, J., S. E. Lynch, J. A. Cribb, S. Kylstra, D. B. Hill and R. Superfine (2018). "Buffer drains and mucus is transported upward in a tilted mucus clearance assay." Am J Physiol Lung Cell Mol Physiol **315**(5): L910-L918.
Chaturvedi, M., M. Kumar and K. Pathak (2011). "A review on mucoadhesive polymer used in nasal drug delivery system." Journal of Advanced Pharmaceutical Technology & Research **2**(4): 215.

Hadjialirezaei, S., G. Picco, R. Beatson, J. Burchell, B. T. Stokke and M. Sletmoen (2017). "Interactions between the breast cancer-associated MUC1 mucins and C-type lectin characterized by optical tweezers." PLoS One **12**(4): e0175323.

Hansson, G. C. (2019). "Mucus and mucins in diseases of the intestinal and respiratory tracts." Journal of Internal Medicine **285**(5): 479-490.

Haugstad, K. E., B. T. Stokke, C. F. Brewer, T. A. Gerken and M. Sletmoen (2015). "Single molecule study of heterotypic interactions between mucins possessing the Tn cancer antigen." Glycobiology **25**(5): 524-534.

Hill, D. B. and B. Button (2012). Establishment of respiratory air-liquid interface cultures and their use in studying mucin production, secretion, and function. Methods Mol Biol. **842:** 245-258.

Hill, D. B., B. Button, M. Rubinstein and R. C. Boucher (2022). "Physiology and pathophysiology of human airway mucus." Physiol Rev **102**(4): 1757-1836.

Kuo, S. C., D. A. Hammer and D. A. Lauffenburger (1997). "Simulation of detachment of specifically bound particles from surfaces by shear flow." Biophysical Journal **73**(1): 517-531.

Kuo, S. C. and D. A. Lauffenburger (1993). "Relationship between receptor/ligand binding affinity and adhesion strength." Biophysical journal **65**(5): 2191-2200.

Lin, J. and M. T. Valentine (2012). "High-force NdFeB-based magnetic tweezers device optimized for microrheology experiments." Rev Sci Instrum **83**(5): 053905.

Mackie, A. R., F. M. Goycoolea, B. Menchicchi, C. M. Caramella, F. Saporito, S. Lee, K. Stephansen, I. S. Chronakis, M. Hiorth, M. Adamczak, M. Waldner, H. M. Nielsen and L. Marcelloni (2017). "Innovative Methods and Applications in Mucoadhesion Research." Macromolecular Bioscience **17**(8): 1600534.

Marshall, B. T., M. Long, J. W. Piper, T. Yago, R. P. McEver and C. Zhu (2003). "Direct observation of catch bonds involving cell-adhesion molecules." Nature **423**(6936): 190-193.

Nathwani, B., W. M. Shih and W. P. Wong (2018). "Force Spectroscopy and Beyond: Innovations and Opportunities." Biophysical Journal **115**(12): 2279-2285.

Niessen, C. M., D. Leckband and A. S. Yap (2011). "Tissue organization by cadherin adhesion molecules: dynamic molecular and cellular mechanisms of morphogenetic regulation." Physiological Reviews **91**(2): 691-731.

Orré, T., O. Rossier and G. Giannone (2019). "The inner life of integrin adhesion sites: From single molecules to functional macromolecular complexes." Experimental Cell Research **379**(2): 235-244.

Sumarokova, M., J. Iturri, A. Weber, M. Maares, C. Keil, H. Haase and J. L. Toca-Herrera (2018). "Influencing the adhesion properties and wettability of mucin protein films by variation of the environmental pH." Scientific Reports **8**(1): 9660.

Wessler, T., A. Chen, S. A. McKinley, R. Cone, M. G. Forest and S. K. Lai (2016). "Using computational modeling to optimize the design of antibodies that trap viruses in mucus." ACS infectious diseases **2**(1): 82-92.

Zanin, M., P. Baviskar, R. Webster and R. Webby (2016). "The Interaction between Respiratory Pathogens and Mucus." Cell Host & Microbe **19**(2): 159-168.

Zhu, M., M. Z. Lerum and W. Chen (2012). "How To Prepare Reproducible, Homogeneous, and Hydrolytically Stable Aminosilane-Derived Layers on Silica." Langmuir **28**(1): 416-423.